# NON-OVERLAPPING COMMUNITY DETECTION

# HOCINE CHERIFI

## Introduction

Many real-world systems are typically made of numerous interacting elements. With simply a local knowledge of the overall system and minimal communications with each other, these individual elements tend to achieve a global goal. Such systems can be adequately described using graph theory. Here, individuals are the nodes of the graph and a link between two nodes represents the interaction. There are many variations of this basic model. For instance, links can be directed, i.e., pointing in one direction. For example, in Web networks nodes are the Web pages, and oriented links represent the hyperlinks pointing to the pages. It is also possible to assign to each link of the graph a weight proportional to the intensity of the connections. To illustrate, in the air-transportation network, nodes represent airports, and there will be a link if there is a flight between two airports. The number of flights between two destinations can weigh the links. In practice, there may also be more than one different type of node or more than one different type of link in a network. A network is said to be "multi relational" or "multiplex" if it contains two or more kinds of relations on a single type of node. This is typical of social networks where the primary objects of study are people and their multiple types of relations. Graphs may contain nodes of distinct types as well. In such cases, they are called "multimodal networks." For example, bipartite graphs contain two types of nodes with links running only from one type to the other. Affiliation networks, in which people are joined together by common membership of groups, can be represented using such a graph. These two types of nodes represent people and groups. A link accounts for the fact that an individual belongs to one group. Besides these extensions, one can also consider many other levels of sophistication such as adding different features to nodes and links and so on.

To understand network behaviour thoroughly, proper knowledge of its topological properties is required. This analysis can be performed at different levels of granularity ranging from microscopic (at the node level) to macroscopic (at the overall network level). However, identifying intermediate scales is a critical issue in order to gain insight into their functional organization. Indeed, communities have always been ubiquitous as elementary forms of organization both in society and nature. As an example, in biology, groups of proteins having the same specific function within the cell can be identified in protein-protein interaction networks. On the World Wide Web, communities correspond to groups of pages dealing with the same or related topics. Therefore, uncovering the community structure is crucial to understanding many real-world networks. It may specifically help to formulate realistic mechanisms for their genesis and evolution.

Although there has been a tremendous effort regarding the community-detection issue, there is no formal consensus on a definition that captures the gist of a community. It is intuitively understood as a cohesive group where members interact with each other more intensely than with those outside the group. As there are many diverse understandings of how cohesiveness translates in formal graph-theoretic terms, various algorithms have appeared in the related literature to discover this hidden structure of a network.

In addition to the variety of definitions, communities can overlap or not. Historically, a great deal of attention has been devoted to non-overlapping community-detection algorithms. In this case, the network is partitioned into separated communities where each node belongs to only one group. Even so, in some situations – particularly in social networks – a node might belong to different communities. In order to reflect this intuition more precisely, there has been a growing interest in the study of overlapping communities in recent years.

Besides the distinction between overlapping and non-overlapping communities, one may consider either global or local approaches to the community-detection problem. Certainly, for networks that are too large and evolve too quickly to have a fully known structure, one cannot rely on global knowledge to uncover a community structure. In order to overcome this limitation, the so called "ego-centred" community methods attempt to find communities related to a single node.

The richness of definitions and features of the community-detection problem has led to an impressive body of literature. In fact, many community-detection methods and surveys have been introduced in recent years. The goal here is to present a state-of-the-art of the most mature research in this area. We will therefore concentrate on non-overlapping community detection with the basic graph model. In this chapter we will give an overview of the most influential approaches to community detection that encompass most of the main methods and techniques. A special focus will also be given to community evaluation.

The chapter is organized as follows: In the preliminaries section, we will review necessary definitions to ensure a good understanding of all elements being addressed. We will then present and analyse the different models used to represent real-world complex networks with a community structure in the network-model section. The taxonomy of community-detection algorithms will be discussed in the following section, and the most influential community-detection methods will subsequently be presented. In the performance evaluation section, we will describe performance measures commonly used to evaluate community-detection algorithms. Finally, test results conducted with synthetic benchmarks will be analysed in order to provide insight into the advantages and drawbacks of the evaluated algorithms

# Preliminaries

In order to understand a complex network fully, one has to analyse its structural features from the microscopic level to the macroscopic level without neglecting the mesoscopic level. The microscopic level concentrates on the differences between individual nodes in order to identify the most influential ones. At the macroscopic level, statistical measures are used to summarise some of the overall network features. Meanwhile, the mesoscopic level results from properties shared by groups of nodes. In this section, we will briefly recall the main measures used to capture, in quantitative terms, their underlying organizing principles.

## Microscopic Topological Properties

The *degree* of a node refers to the number of links attached to the node. It can be understood as a measure of the node's leadership in the network. Highly connected nodes, referred to as hubs, are critical elements of the networks. A *bottleneck* is a node controlling the flow from one part to another part of the network whose deletion increases the number of unconnected sub-networks. A *bridge* is a link connected to bottlenecks. As such, the network is separated into unconnected sub-networks when the bridges are removed.

*A clique* in an undirected graph is a subset of nodes set such that each node is linked to all the other nodes of the subset (complete graph). Its size corresponds to the number of nodes it contains.

*The local clustering coefficient* quantifies the embeddedness of a node in a clique. It is ascertained by taking the proportion of links between the nodes within its neighbourhood divided by the number of links that could exist between them. Its value is 0 if none of the node's neighbours are connected, and its value is 1 if all of the neighbours are connected. The origins of the local-clustering coefficient can be traced back to sociology, where similar concepts have been used. In a typical social network, a person's friends are very likely to know each other. This inherent tendency for people to cluster in circles of friends, in which every member knows every other member, is quantified by the clustering or transitivity coefficient. Such a concept – known also as a "fraction of transitive triples" in sociology – is used to capture the degree of social embeddedness that characterizes the nodes.

*The distance between two nodes* is the number of links in the shortest path existing between those nodes in which each link and node appear at most once.

*Centrality* determines how influential a node is within a network. Degree, closeness, and betweenness centrality are the most widely used measures to define such a characteristic. *Degree centrality* measures the involvement of a node in the network by the number of nodes connected to it. This local definition does not take into account the position of the node in the network and therefore cannot measure its ability to reach others quickly. *Closeness centrality* captures this feature. It is based on the inverse sum of the shortest distances to the other nodes of the network. *Betweenness centrality* asserts the ability of a node to play a "broker" role in the network by measuring how well it lies on the shortest paths connecting other nodes.

## Macroscopic Topological Properties

Undirected real-world networks are known to share some common properties. In this section, we will present the most usual properties that are implemented to characterize complex networks.

*Density* is the ratio of the number of existing links to the number of possible links. For a complete network (where all nodes are connected to all other nodes), the density is 1. A *dense network* is a network in which the number of links of each node is close to the number of nodes. A sparse network, by contrast, is connected by a low number of links only.

*The small-world* property refers to the low average distance value between any two nodes of a network. This property has been popularized by the ''six degrees of separation'' concept, i.e., anyone on the planet can be connected to any other person through a chain of acquaintances that has no more than five intermediaries on average. This is typical of many real-world networks where shortcuts connecting different areas of the networks allow for the reduction of distance between any two network nodes. For instance, on the Internet, a computer can be reached on average through six routers, and the co-authors in mathematics are on average within four authors from each other.

*The global clustering coefficient* reflects the tendency of link formation between neighbouring nodes in a network. It is also called *transitivity*. This characteristic can be measured by the mean value of the local clustering coefficients. (Note that alternative definitions exist for this quantity as well.) Transitivity is known to be higher in real-world networks as compared to uncorrelated random networks in which triangles are sparse.

*Degree distribution* measures the statistical repartition of the network nodes' degrees. Along these lines, one of the most interesting developments in our understanding of complex networks is that they exhibit an inhomogeneous distribution with few nodes linked to many other nodes and a large number of poorly connected nodes. In particular, for a large number of networks, such distribution can be adequately described as a *power-law distribution*. These networks are often referred as "*scale-free networks*" because their degree distribution does not depend on their size. Related experimental studies show that the exponent value of the power law usually ranges from 2 to 3 (Albert and Barabasi 2002; Boccaletti et al. 2006; Newman 2003).

*The degree correlation* measures the tendency of nodes to associate with other nodes sharing the same characteristics and especially the same degree values. In *assortative* networks, the nodes tend to associate with their connectivity peers, and the degree correlation is positive. In *disassortative* networks, high-degree nodes tend to associate with low-degree ones, and the degree correlation is negative. Social networks appear to be assortative while information, technological and biological networks appear to be disassortative.

*The centralization* measures the ability of a network to be focused around a few central nodes. Centralization measures are based on the differences between the centrality scores of the most central point and those of all other points. The definition of centralization measures is general, so it can be based on any centrality measure. Its value ranges from 0 to 1 for the local centrality concepts presented previously. Correspondingly, a value of 0 is obtained on all three measures for a "complete" graph, while a value of 1 is achieved for a "star" or "wheel" graph. Very centralized networks are very sensitive to the dominating central node failures or attacks, while less centralized ones are more resilient.

## Mesoscopic Topological Properties

At this scale, the communities are elementary units of the topology. Before addressing community properties, the first issue to be solved is to express a formal definition of a community. This is quite a challenging task because until now, the debate has remained open so that there is no consensual definition. There are also many interpretations of the intuition of cohesiveness and separation tied to the community notion. Moreover, in many cases, community-detection algorithms are not based on a formal definition of the communities. In an exhaustive survey, Fortunato (2010) identifies three types of definitions: local, global, and similarity-based definitions.

*Local definitions* consider communities as autonomous entities that can be evaluated independently of the rest of the graph except for their local neighbourhood. Cohesion is related to basic patterns such as a clique or relaxed variants. The community structure is obtained by searching for maximal subsets of the pattern.

In *global definitions*, communities are defined with respect to the graph as a whole, and a global criterion is used to uncover the community structure. More often, the criterion compares a property of the graph to a similar random graph known to have no community structure. Many global definitions are based on the density of nodes guided by the intuition that a community is a group of densely connected nodes that are sparsely connected to the rest of the network. Community-detection algorithms try to maximise the number of links inside the groups while minimising the number of edges between nodes in different groups.

*Node-similarity definitions* assume that communities are made of nodes with common properties. Irrespective of whether they are connected by an edge or not, nodes are classified according to a similarity measure of a given local or global property (distance, commute time between nodes, etc.). Recently, Coscia et al. (Coscia, Giannotti, and Pedreschi 2011) introduced a meta-definition of a community based on the following intuition: a community is made of entities that are closer to others within the community than to the ones outside it. A community is defined as a set of entities sharing a "common set of actions". This meta-definition encompasses the vast majority of models used in the related literature. For example, considering the links between two nodes as a particular kind of action allows for representing density-based definitions. Node-similarity-based definitions can be apprehended by modelling the similarity measures as the similarity of the action set. However, note that some definitions of community cannot fit into these taxonomies.

While the extreme richness of definitions and features has led to the publication of extensive literature on the community-discovery problem, few authors have been interested in the analysis of real-world network community structures. An important issue in this regard is whether or not there is a characteristic size for a community. Studies show that *community-size distribution* is inhomogeneous. It can be adequately described by a power law (Guimerà et al. 2003; Newman 2004a) with an exponent ranging from 1 to 3 (Palla et al. 2005). Thus, many small communities coexist with a few very large ones.

The "*embeddedness*" of a node measures the proportion of its neighbours belonging to its community. Its value is 1 when all of the node neighbours are in its community, whereas it is 0 when all its neighbours belong to different communities. In real-world networks, low-degree nodes predominantly have a very high embeddedness. Some networks (like the Internet, communication networks and biological networks) exhibit a peak of around 0.5, whereas others have a more uniform distribution (such as social networks and information networks) (Lancichinetti et al. 2010).

The *density* of a community indicates the cohesion of the community as compared to the overall network density. Undoubtedly, a community is supposed to be denser than the network it belongs to. The *scaled density* is obtained by multiplying the density by the community size (Lancichinetti et al. 2010). Its value is 2 when the community has a tree structure. On the other hand, if it is a completely connected subgraph, its value is the size of the community. According to this measure, real-world networks such as the Internet or communication networks have essentially tree-like communities. Meanwhile, on social and information networks, the scaled density increases with the community size, suggesting a clique-type behaviour. Then we have biological networks, which exhibit a hybrid behaviour as their small communities are tree-like, while the large ones are denser and close to cliques (Lancichinetti et al. 2010).

*Hub dominance* reveals the presence of hubs in communities. This factor corresponds to the maximal internal-degree ratio in the community and the maximal theoretical degree given the community size. Its value is 1 when at least one node is connected to all other nodes in the community. Yet it can reach 0 in the unlikely case where no nodes are connected. For communication networks, the hub dominance value is close to 1 and rather independent of the community size. More generally, complex networks do not have as many hubs in their large communities. This is why their hub dominance generally decreases when the community size increases (Lancichinetti et al. 2010).

*Modularity* measures the deviation of "intra-community" links' repartition of a given community structure compared to the one obtained in a similar network with links placed at random. This measure is based on the idea that the number of links between nodes in a community is higher than in a random network. It has been introduced by Newman and Girvan to measure the quality of a partition. When the communities are not better than a random partition, or when the network does not exhibit any community structure, its value is low. Moreover, it can be negative if the network has a disassortative structure. For real-world networks, modularity values between 0.3 and 0.7 are considered high. Modularity is used as a performance measure to quantify how good a given network partition is. It is also an optimisation criterion in many community-discovery algorithms. Nonetheless, one should note that modularity presents some drawbacks. For example, it is sensitive to community size (Fortunato and Barthélemy 2007; Lancichinetti and Fortunato 2011) and it is possible to find partitions of random networks with relatively high modularity values (Guimera, Sales-Pardo, and Amaral 2004).

# Networks Models

Traditionally, large networks with no apparent design principles have been described as "random graphs." The classical Erdös-Reyni random network model, abbreviated as ER, is the most influential model to represent such graphs. It is based on the assumption of independent and purely random link formation. Even if it shares with real-world networks a short average distance, this model does not capture the main characteristics of real-world data. Indeed, the clustering coefficient is several orders of magnitude larger in real-world networks. Furthermore, its degree distribution follows a Poisson law, while a power law is a better fit for real-world networks. Note that the power law indicates the presence of hubs, whereas these highly connected nodes have a very low probability in Poissonian distribution. Given that the Erdös-Reyni model is evidently unable to mimic real networks, various attempts have been made to find appropriate theoretical models. In addition, since the seminal paper of Watts and Strogatz on "small-world networks" was published, activity in this field has been growing. From that time on, the major focus of research has moved from small-world networks to "scale-free" networks. Within this framework, there exist two main approaches: static or dynamic network models. Static models give a snapshot of a real-world network, while dynamic models try to capture how the network actually grows.

The so called "inhomogeneous random graph" is a static model aimed at generating graphs with a given degree distribution. To do so, equal edge probabilities of the Erdös-Reyni random graph are replaced by edge-occupation statuses that are independent and are moderated by certain node weights. The Configuration Model, abbreviated as CM, is another static model introduced to generate random networks with controlled degree distribution (Molloy and Reed 1995). It operates in two steps. First, to each node a number of half links called "stubs" is assigned at random according to the degree distribution. Then, pairs of stubs are chosen at random, and a link is made between the corresponding nodes. So even though the node degree distribution of the graph remains intact, the configuration model results in a completely random network. The reader should observe that vertices having self loops as well as multiple edges may occur. However, such types of connections are scarce. Furthermore, one can add a constraint to avoid such patterns.

The network models discussed thus far assume that the size of the graph is fixed, while many real-world networks grow continuously throughout the lifetime of the network by the addition of new nodes. In order to explain the network-formation process, Barabási and Albert (1999) popularized the concept of preferential attachment introduced by Yule in the context of species evolution. Their model, hereafter referred to as the BA model, is based on two ingredients: growth and preferential attachment. Starting with a small network of arbitrary structures where each node is at least connected to one other node, new nodes are sequentially added with a number of links connected to them. These links are attached to another node with a probability proportional to the degree of the receiving node. Such a process, also known as "the rich get richer," favours highly connected nodes. For a large network size, it leads to a scale-free degree distribution with an exponent value equal to 3. Following the publication of Barabási's and Albert's study, a large body of work investigating dynamic models appeared in the relevant literature. Building on this model, many fitness-based models have been proposed (Nguyen and Tran 2011). The basic idea of such models is to use more sophisticated features than the simple degree value to explain the propensity of a node to attract new links. Once a fitness value is associated with each node, the growing process can be implemented.

To date, many network models can generate scale-free features based on different ideas and mechanisms, but few network models with community structure have been proposed. In the rest of this section, we provide an overview of the most famous benchmarks specifically designed for community-structured complex networks. Also, we report a recent work showing that these models can be improved to reproduce the structural properties observed in real networks.

In order to test the performance of community-detection algorithms on networks with varying degrees of community structure, Girvan and Newman (2002) introduced one of the first models (abbreviated as GN). Each graph is constructed with 128 nodes separated in 4 communities of equal size and each node is linked on average to 16 nodes. Links joining nodes inside a community are placed independently at random according to a given intra-community probability value. Conversely, nodes of different communities are linked at random according to an inter-community probability value. The strength of community associations is controlled by the ratio of intra-community to inter-community links. Although this model deviates drastically with measurements from real-world networks, it has been widely used in order to test community-detection algorithms (Donetti and Munoz 2004; Duch and Arenas 2005;

Girvan and Newman 2002; Radicchi et al. 2004). First of all, the network size is not realistic. Nowadays, many complex networks are actually made of millions of nodes. Furthermore, the nodes' degrees follow a Poisson distribution, while in many real networks, the degree distribution displays a fat tail. Finally, communities have identical size, whereas experimental studies on real networks show that the community-size distribution follows a power law. In order to overcome these limitations, several variants of this model have been defined producing larger networks and communities with heterogeneous sizes (Danon, Diaz-Guilera, and Arenas 2006; Pons and Latapy 2005).

More recently, Bagrow (2008) introduced a different approach based on rewiring an initial network such as a community structure. Starting from a graph with a given degree distribution, nodes are randomly assigned to different communities. Afterwards, a random pair of inter-community links is rewired to be intra-community links. This switching process of the link-pair extremities preserves the degree distribution of the network. In his experiments, Bagrow used the BA model to generate the initial network. Meanwhile, the community structure is made of four equally sized communities, and 25% of the edges are rewired. Keep in mind that since the partition is random, the initial modularity is very small. This is because with edges being moved within communities, the modularity increases with the proportion of rewired pairs.

Lancichinetti, Fortunato and Radicchi (2008) introduced a model based on a rewiring principle, which has obvious advantages over Bagrow's scheme. In this model, denoted as LFR, users chooses the network size. Furthermore, the degree of the nodes and the community size can be adjusted in order to follow a power-law distribution while controlling the exponent value. The LFR algorithm produces networks with non-overlapping communities using a two-step process. First, a scale-free network with controlled size $n$ is generated using the configuration model. The minimum and maximum degree are chosen such that the degree distribution follows a power law with a given average degree value $<k>$ and exponent $\gamma$. Then, the second step is applied in two phases. First, the communities are randomly drawn so that their distribution size follows a power law with a given exponent value $\beta$. Second, an iterative rewiring process takes place to control the fraction of links $\mu$ shared by a node with nodes in other communities. As it is generally not possible to meet such a constraint exactly, this tunable parameter called the "mixing coefficient" is only approximated in practice. Its value determines how clearly the communities are defined. For small $\mu$ values, the communities share only a few links and are therefore clearly separated. Increasing $\mu$, the proportion of inter-community links grows and the original communities gradually disappear.

The configuration model used in the first step of the LFR model is very flexible as it is able to produce networks with any size or degree distribution. Nevertheless, it is known to generate networks with no correlation between the degrees of connected nodes (Serrano and Boguñá 2005) and low transitivity when degrees are power-law distributed (Newman 2003). To overcome these drawbacks, more realistic models can be used. For this purpose, in (Orman, Labatut, and Cherifi 2013) two alternative models have been evaluated : the BA model and a fitness-based model called *evolutionary preferential attachment* (Poncela et al. 2008) abbreviated as EV. Both models generate scale-free networks with a desirable size and average degree.

In the BA model, the capacity of a node to attract links from newcomers depends on its current degree value while in the EV model it is linked to its evolutionary fitness. This dynamical-variable value is proportional to the payoffs obtained when playing a prisoner's dilemma game with its neighbours. In this setting, nodes with higher scores are more attractive to newcomers because their strategy has proven to be successful. Meanwhile, in every iteration, each node plays either cooperation or defection against all of its neighbours. It gets a total score depending on the individual results: 0 for unilateral cooperation or bilateral defection, 1 for bilateral cooperation, and $b$ for unilateral defection, with $b > 1$. The first move is randomly chosen, whereas the next one depends on the respective results of the considered node and a randomly picked neighbour. If the neighbour's score is better, the node might switch its strategy with a probability depending on the difference between their scores. A parameter called the "selection pressure" ($\varepsilon$) allows modulating the influence of fitness in the attachment mechanism. The evolutionary dynamic has no influence for $\varepsilon = 0$ while the fitness scores are fully considered for $\varepsilon = 1$. According to its authors, when the selection pressure is large, networks generated by this model display both high transitivity and degree-correlation values.

Hereafter, we will refer to the network models by using the following abbreviations: LFR-CM denotes the original LFR method, while LFR-BA and LFR-EV are the modified versions utilising respectively BA and EV in the first step of the algorithm. In order to illustrate the comparative realism of the networks generated by the three algorithms, we report representative results on the degree correlation, the clustering coefficient, and the degree centralization on figure 11-1. More details are given in (Orman, Labatut, and Cherifi 2013). For every value of the mixing parameters ranging from 0.05 to 0.95 with a step value of 0.05, 25 networks have been generated, and the mean and standard deviation values of the topological properties have been computed. Note that parameters values have been chosen such that generated networks exhibit realistic topological properties.

The models exhibit a very different behaviour concerning the degree correlation. Specifically, for LFR-EV, the degree correlation decreases almost linearly when the proportion of inter-community links increases. Whatever the mixing coefficient value, it exhibits a rather higher degree correlation value as compared to its competitors. In contrast, the degree correlation increases with the mixing parameter in LFR-BA. This almost linear growth is nevertheless observed with a lower slope and lower variance estimates. We also have LFR-CM, which is the less realistic benchmark. Starting from a reasonable value when the communities are well separated, it decreases rapidly and oscillates around zero for mixing coefficient values greater than 0.4.

The behaviour of the three algorithms is quite similar regarding the clustering coefficient. As the proportion of inter-community links increases, transitivity decreases. Nonetheless, the three models exhibit realistic transitivity when the communities are more or less well separated. Most definitely, real-world networks with a transitivity greater than 0.3 are considered highly transitive (da F. Costa et al. 2008). Whatever the mixing coefficient value, LFR-BA is always the worst, while the results are more mixed for the two other models. In addition, LFR-CM outperforms LFR-EV when the mixing coefficient is less than 0.3, and the latter takes a slight advantage for higher values. In any case, the results are not very satisfactory for high $\mu$ values.

Degree centralization is very insensitive to the mixing-proportion variations. It is higher for LFR-BA and LFR-EV as compared to LFR-CM. Moreover, there are virtually no influential nodes in the networks generated by LFR-CM as suggested by the very low values of the centralization. The attachment mechanism used in the dynamic models tends to generate nodes highly connected to their neighbours. As such, the presence of these hubs results in higher centralization values. Note that the same behaviour is observed with closeness and betweenness centralization.

Overall, we can conclude that dynamic models produce more realistic networks, at least for the degree correlation and the centralization as compared to the original LFR model. Even so, except for well separated communities, the three models fail to reproduce networks with high transitivity values. The explanation may originate from the rewiring process or more fundamentally from the construction process of these networks. Indeed, pairwise relations (dyads) between nodes make for the fundamental unit when modelling a network, regardless of whether they come by growth or probabilistic models. The basic assumption underlying these models is that dyads are independent. However, the large clustering coefficient observed in many networks hints at dependence between the connections. This observation thusly suggests that models should be defined based on triads (sets of three nodes) rather than dyads (Winkler and Reichardt 2013).

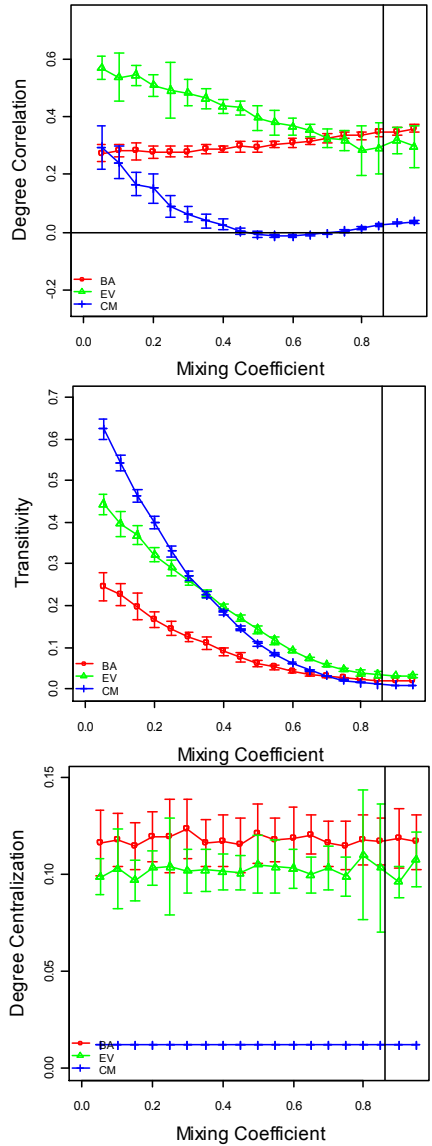

**Fig. 11-1.** Variation of macroscopic topological properties with the mixing coefficient μ. Red squares represent LFR-BA, green triangles LFR-EV, and blue crosses LFR-CM. The vertical lines represent the average mixing limit above which communities stop being clearly defined. Networks parameters are $n$ = 5000, $<k>$= 30, γ ≈ 3, β=2.

# Algorithms

Community detection is a prolific subject in the literature regarding complex networks. A great variety of algorithms have been developed so far to deal with this issue. However, there has evidently been no clear and precise definition of what a community is, so the problem has been dealt with from many points of view. For instance, a community has been expressed as a graph partitioning, community mining, spectral analysis, an optimisation problem, a statistical problem, and so on. Some recent surveys that aimed at providing an overview of the field have proposed taxonomies of the community-detection methods. Even in the relatively mature area of non-overlapping community detection, there is no consensus on a classification. To this end, the main difficulties lie in the fact that there is an overlap of methods and in the underlying definition used, if any. Finally, work is still in progress, and new solutions are constantly scrambling the proposed taxonomies.

Not to add to the confusion, we adopt in the following section the classification proposed in (Coscia, Giannotti, and Pedreschi 2011). To our knowledge, this is the most recent work on the subject. Furthermore, rather than focusing on how communities are detected, the classification is based on the definition of the community used by the algorithms. Starting from a meta definition of a community, algorithms are classified into eight categories according to different interpretations of the meta definition. Using the same terminology, we present the most influential categories. Among the various solutions existing in the pertinent literature, we also present a representative sample of community-detection algorithms known for their effectiveness or their popularity.

## Internal Density

Internal density is highly influential in the community-detection literature. It is based on the widespread assumption that a community is a set of densely connected nodes. Generally, the algorithms in this category try to expand or collapse the node partitions in order to optimize a given density function. Despite the fact that different density functions have been proposed, the key concept linked to the approach is modularity. Under this definition, a community is a set of nodes with a higher density than what would be expected in a random network with the same degree distribution. Numerous solutions to this NP complete-optimisation problem have been proposed ranging from greedy heuristics to stochastic formulations. We present here three widespread algorithms.

*FastGreedy* is an agglomerative hierarchical clustering method developed by Newman et al. (Newman 2004b). Starting with each node in its own community, it iteratively merges the two communities producing the greatest modularity increase. At each step, communities sharing one or more links are allowed to merge. The process stops when all of the nodes are in the same final community. At this point, the outcome is a dendrogram representing the hierarchical decomposition of the network into communities. The optimal community structure is obtained by cutting the dendogram at the maximum point of its modularity.

*Louvain* is another greedy agglomerative hierarchical algorithm proposed by Blondel et al. (2008). In this case, two phases are repeated iteratively. Starting with each node in its own community, the gains in modularity obtained by placing a node in the same community than its neighbours are evaluated. The community offering maximal gain is then retained. This process is applied repeatedly and sequentially for all nodes until no individual move can improve the modularity. At the end of this first phase, the algorithm yields the first partitioning scheme. In the second phase, a new network is built whose nodes are the communities found during the first phase. The intra-community links are represented by self-loops, whereas the inter-community links are aggregated and represented as links between the new nodes. It is then possible to start the first phase again on the resulting weighted network. This process is iterated until only one community remains.

*Spinglass* was introduced by Reichardt and Bornholdt (Reichardt and Bornholdt 2006). The authors show that searching for the optimal modularity partition is analogous to finding the spin configuration that minimizes the energy of a statistical physical model named *Potts spin glass*. In this setting, the spin states are the hidden community membership. Simulated annealing is used to minimise the non-quadratic energy function. Note that any other scheme that can deal with this combinatorial optimisation problem can be implemented.

*Leading Eigenvector* is a spectral partitioning technique introduced by Newman (Newman 2006). While in traditional methods, a good partition is one that minimises the number of links running between the communities (known as cut size), Newman claims that a good partition is one where it is smaller than expected. In other words, rather than minimizing the cut size, he proposes to minimise the modularity. To do so, the modularity is written in matrix form, and the principal component analysis is applied on this matrix.

Notice that although widely used, modularity suffers from some limitations. In particular, it fails to identify communities smaller than a network-dependent critical size. For a good review of modularity limitations, refer to (Good, De Montjoye, and Clauset 2010).

## Closeness

The closeness approach is based on the assumption that a community member can reach any other member of the same community by crossing a smaller number of links as compared to the network's average distance. Undoubtedly, a community is supposed to be a smaller world than the overall network. Inspired by this definition, several algorithms rely on random walks to uncover the community structure. Recall that in a random walk, a walker performs a series of sequential moves from node to node, and each move is chosen randomly from its neighbours. The basic idea of these algorithms is to perform several random walks and to cluster the nodes frequently crossed in the same walk. We will present three well-known algorithms of this type.

At the heart of the *Markov Cluster* algorithm lies the idea of simulating a controlled flow through random walks in order to detect communities (Van Dongen 2008). The basic idea here is that the current is strong *in* the communities yet weak *between* communities. Promoting the flow where the current is strong and mitigating the flow where the current is weak will result in no flow between communities, thus revealing the community structure of the network. This algorithm simulates random walks by iteratively applying two transformations (expansion and inflation) on the network-transfer matrix until convergence occurs. The final matrix can be interpreted as the adjacency matrix of a network with disconnected components representing the communities. Expansion corresponds to the computation of random walks of higher length. Since higher length paths are more common within communities than between different communities, the probabilities associated with node pairs lying in the same cluster will, in general, be relatively large as there are many ways of going from one to the other. Inflation changes the probabilities of random walks, favouring more probable walks over less probable ones. This operation boosts the probabilities of intra-community walks and demotes inter-community walks. Eventually, iterating expansion and inflation results in the separation of the graph into distinct components interpreted as communities.

*WalkTrap* (Pons and Latapy 2005) is based on the intuition that random walks are able to unveil the real distance among nodes. Based on the information given by random walks, a distance that captures structural similarity between the nodes is computed. At this point, the community-detection problem reduces to a clustering one. The algorithm thusly uses a hierarchical agglomerative approach. At each step, the two communities that minimize the variation of the squared distances' mean value between each node and its community are merged. Ultimately, to find the optimal community structure, the dendrogram is cut at the maximum point of the modularity.

*InfoMap* (Rosvall and Bergstrom 2008) uses a compression technique to describe the information flow on networks. Random walks of a given length and with a given probability of jumping to a random node are performed. Each walk is described as a sequence of steps inside a community followed by a jump through a two-level nomenclature based on Huffman coding. The first one is used to distinguish communities in the network and the other to distinguish nodes in a community. Each node codeword is derived from the visit-node frequency of an infinitely long random walk. This coding strategy leads to a compact representation of the walks. Indeed, with a partition featuring limited inter-community links, the walker is statistically more likely to stay longer inside communities. Therefore, only the second part of the nomenclature is needed to describe its path. The authors showed that the optimal partitioning problem is reduced to finding the minimum description length for all of the walks.

## Diffusion

The main idea with diffusion is to consider a community as a set of nodes grouped by the propagation of the same property or information in the network. Algorithms based on this definition usually perform a diffusion on the network following a particular set of transmission rules and then group together any nodes that end up in the same state.

*Label Propagation* (Raghavan, Albert, and Kumara 2007), simulates the diffusion of some information in the network to identify communities. Initially, each node carries a label denoting the community to which it belongs. Then, each node updates its community based on the labels of its neighbours. The rule used is to join the most frequent community in its neighbourhood (ties are broken randomly). As the labels propagate, densely connected groups of nodes reach a consensus on a unique label. This process goes on until each node has the majority label of its neighbours.

Even if we focus on only one algorithm of this class, it is worth mentioning influence-spreading propagation techniques. The idea is to take advantage of the influence propagation in order to identify a group of users – often called tribes – that behave homogeneously (Chen, Wang, and Yang 2009; Goyal, Bonchi, and Lakshmanan 2008; Khorasgani, Chen, and Zaäne 2010).

## Bridge Detection

The bridge-detection approach relies on the idea that a community is made of dense components linked by a few bridges. Removing these bridges allows for uncovering the communities. The core issue is to define these bridges, which can be either nodes or links. The basic procedure is to compute a measure of either nodes' or links' contribution to keep the network connected and to remove the ones with the highest scores. We will present two methods. The first one is based on a global-link-centrality measure, while the second utilises a local-link-centrality definition.

The most popular algorithm in this category, *Edge Betweeness,* uses a link centrality measure. Proposed by Newman (Girvan and Newman 2002), *Edge Betweeness* measures the centrality of a link by considering the proportion of shortest paths going through it. The main assumption here is that all of the shortest paths between different communities must go along the bridges leading to high scores for these links. The algorithm uses a divisive hierarchical approach. Starting with the initial network, the most central link is removed, and the link centrality is computed on the remaining network. When ties exist, one edge is to be chosen at random. Accordingly, this process is iterated until no links remain. The main drawback with this algorithm is that centrality must be computed at each step, making it unsuitable for large networks.

*Radetal*, (Radicchi et al. 2004) uses an alternative centrality measure. This local measure is defined in analogy with the usual node-clustering coefficient, as the number of triangles to which a given edge belongs, divided by the number of triangles that might potentially include it. Note that, unlike betweenness centrality, this measure called link clustering takes high values for links inside a community and low ones for bridges. Indeed, clustering is higher in a community while links connecting nodes in different communities are included in few or no triangles. Radetal works exactly as Edge Betweeness with the difference that, at every step, the removed edges are those with the smallest edge clustering value.

# Evaluation Criteria

Objective evaluation of the community-detection algorithms is a strategic issue. Indeed, we need to verify that the communities identified are actually the good ones. Moreover, it is necessary to compare results between two distinct algorithms to determine which is most effective. This complex and open problem is mainly considered from the evaluation-of-clustering-algorithms perspective. In such a setting, validation is simply accomplished by comparing discovered communities against known ones. Various clustering-comparison measures have been proposed that can be classified into three main categories: measures based on pair counting, set-matching-based measures and information-theoretic-based measures. Because of the overwhelming number of measures and their heterogeneity, choosing the most adapted is a difficult problem. The following section presents the standard evaluation measures emerging for the evaluation of community-detection methods. Note that all of the measures presented here can be derived from the confusion matrix whose elements are the number of nodes that are common to both partitions.

## Pair-Counting-Based Measures

With pair-counting-based measures, clustering comparison is based on counting the pairs of points on which two partitions agree or disagree. Any pair of points can be classified under one of the following four categories. In the first two cases, the two partitions are in accordance: either the pair of points is in the same cluster in both partitions or it does *not* belong to the same cluster in both partitions. In the two remaining cases, the partitions disagree. It is when a pair of points belongs to different clusters in a partition but is in the same cluster in the other partition. The *Rand Index* (Rand 1971) is a well-known measure in this class. It computes the proportion of agreement between the two partitions. Its value is 1 when the two partitions are identical and 0 when no pair of points appears either in the same cluster or in different clusters in both partitions. This extreme situation happens only when a partition consists of a single cluster. Meanwhile, the other consists only of clusters containing single points. As the expected value between two random partitions does not take a constant value, Hubert and Arabie (Hubert and Arabie 1985) proposed a corrected for chance version of the Rand Index. The so called *Adjusted Rand Index* takes on the value 0 when the two partitions are picked at random. Negative values indicate a strong divergence between the partitions. The *Jaccard Index* is the ratio of the number of point pairs classified in the same cluster in both partitions, to the number of pairs classified in the same cluster by at least one partition. There is a corrected for chance version of this measure. It should be remembered that, there are many other measures in this class (Albatineh, Niewiadomska-Bugaj, and Mihalko 2006). However, after correction for chance, many of these measures are equivalent (Warrens 2008).

## Set-Matching-Based Measures

Set-matching-based measures are based on set cardinality. They intend to find the largest overlaps between pairs of different partition clusters. *Purity* is the proportion of correctly assigned nodes. Each identified cluster is matched to the one with the maximum overlap in the reference cluster, and then the accuracy of this assignment is measured by counting the number of correctly assigned nodes. When community structures are very dissimilar, the purity value is close to 0, while a perfect clustering has a purity of 1. High purity is easy to achieve when the number of clusters is large. In particular, purity is 1 if each node gets its own cluster. The *classification error* (Meila 2005) and the *Van Dongen metric* are two alternative measures aimed at finding the best match for each cluster in both partitions. The main drawback of these measures is that they fail to take into account some clusters when their overlap with the other partitions is not enough large.

## Information-Theoretic-Based Measures

Information-theoretic-based measures have gained increasing attention in the clustering literature. They are based on the *mutual information* shared by two partitions in order to assess their agreement. When two partitions are independent, they do not share any information. On the contrary, when they have the same distribution, the information shared is maximum. Knowing one of the partitions gives a perfect knowledge of the other one. This mutual information is a non-negative quantity upper bounded by the entropies of both partitions. We also have the *variation of information* that was introduced by Meila (2005). This metric is defined as the sum of the entropies of two partitions minus two times the mutual information. It measures the amount of information lost and gained in changing from one partition to the other. When the two partitions are identical, its value is 0, and its upper bound depends on the size of the network. *Normalized mutual information* (NMI) is defined as the ratio of the mutual information to the mean value of the entropy of both partitions. It takes the value of 1 when the two partitions are identical and 0 when they are independent.

## Quality Functions

When there is no ground truth, the only way to evaluate a partition is to use a quality function. Existing quality functions formalise in different ways the idea that communities are sets of nodes densely connected and poorly connected to the rest of the network. In (Yang and Leskovec 2012), eleven quality functions are reported. The authors show that these functions can be grouped into four classes: quality functions based on internal connectivity of the partition, quality functions based on external connectivity of the partition, quality functions based on internal and external connectivity, and quality functions based on a model. In the first group, we notice the *internal density*, which is the internal-link density of the node set. In the second group, the *cut ratio* is the fraction of existing edges out of all possible links leaving the communities. *Conductance* is a prominent measure of the third group. It measures a fraction of the volume of the total edges that points outside the cluster. The fourth group includes the modularity and its variants. *Modularity* is the most widely used quality function to compare the effectiveness of community-detection algorithms on real data when the underlying community structure is not known. This approach expresses how a community structure has a high-density ratio as compared to a random graph with the same degree sequence. The main weakness of the modularity approach is that it is also an optimization criterion used by a large number of algorithms; using it as a quality function introduces a bias in the comparisons. Furthermore, it has been reported that a community-detection method yielding strong modularity results is not always the best choice.

More recently, an alternative to modularity called *surprise* has been proposed (Aldecoa and Marín 2013). It assumes as a null model that links between nodes emerge randomly. It then evaluates the departure of the observed partition of the expected distribution of nodes and links into
communities according to the null model. The higher the surprise value the
more unlikely the partition is a realisation of a random graph.

# Tests on Synthetic Networks

The evaluation of community-detection algorithms is a complex
problem that has received very little attention in the related literature. We
can distinguish two main trends for assessing the performance of
community-detection methods based respectively on subjective or
objective evaluations. Subjective evaluation relies on a panel of experts
who decide whether or not the community structure revealed by the
algorithm is valid. This decision depends on the relevance of results in the
domain. In the subjective procedure, which is difficult to carry out on large
networks, the results may vary depending on the expert and on the angle of
interpretation adopted. On the other hand, objective evaluation is
conducted using a set of benchmark graphs with a well known community
structure and one or more evaluation measures chosen among the ones
presented in the previous section. Both real-world and synthetic data can
be used for this purpose. In some real-world networks, group membership
for the nodes is explicitly defined. It is used to define ground-truth
communities. A typical example, called the *Zachary's Karate Club*
network, is widely used in the literature. This small graph is made of two
groups that are easily recovered by any community-detection method. The
*Stanford Large Network Dataset Collection*[1] is a solid attempt at
establishing a standard real-world benchmark. It contains a set of social,
information and collaboration networks with ground-truth communities.
Networks in this benchmark cover a wide range of density, size and
community structures. However, as of now, there has been no consensus in
the scientific community on a standard real-world data set to explore the
properties of community-detection algorithms. Most papers introducing a
new method just use the simple classical benchmark graphs that can be
visually interpreted in order to quantify the performance of their algorithm
on real data. Things evolve more favourably regarding synthetic data. To
date, the LFR benchmark is a de facto standard used for the large-scale
evaluation of diverse community-detection methods. In what follows, we
will focus on the objective assessment of artificial benchmarks. In this
case the networks can be arbitrarily designed and controlled to highlight
the strengths and weaknesses of community-detection algorithms in a
broad range of situations. We will briefly review recent works on
experimental evaluation, concentrating on the algorithms presented above,
with two questions in mind: 1) Is there a community definition better
suited than another according to performances? 2) How do the performances evolve according to the degree of the benchmarks' realism used in the evaluation?

---

[1] http://snap.stanford.edu/data/

Table 11-1 reports the results of an experimental evaluation of the algorithms presented in the previous section using the LFR model with 5,000 nodes. The exponent values of the power law of the degree distribution and of the community-size distribution are respectively $\gamma \approx 3$ and $\beta = 2$. The average degree is $<k> \approx 30$, and the maximum degree is $k_{max} = 90$. Two different values of the mixing parameter $\mu$ {0.2, 0.6} are used. The first one corresponds to well separated communities, while the second one is not far from the limit where the networks have no community structure. For each parameter set, 25 networks are generated. The Rand Index, the Adjusted Rand Index and Normalized Mutual Information have been computed to assess the performance of the algorithms. For clarity, only the mean NMI values are reported.

| Algorithm | NMI | Rank | NMI | Rank | Mean Rank |
|---|---|---|---|---|---|
| Mixing Proportion | $\mu=0.6$ | | $\mu=0.2$ | | |
| InfoMap | 1.0 | 1 | 1.0 | 1 | 1 |
| WalkTrap | 0.98 | 2 | 1.0 | 2 | 2 |
| Label Propagation | 0.98 | 2 | 1.0 | 3 | 2,5 |
| Spinglass | 0.95 | 4 | 0.97 | 4 | 4 |
| Louvain | 0.93 | 5 | 0.95 | 5 | 5 |
| Markov Cluster | 0.78 | 5 | 0.92 | 6 | 5,5 |
| Radetal | 0.78 | 5 | 0.80 | 7 | 6 |
| FastGreedy | 0.41 | 8 | 0.48 | 8 | 8 |
| Leading Eigenvector | 0.28 | 9 | 0.37 | 9 | 9 |

**Table 11-1.** Performance evaluation of community-detection algorithms for well separated communities ($\mu=0.2$) and mixed communities ($\mu=0.6$)

The result of this empirical evaluation illustrates the general behaviour of the algorithms well. It clearly shows that algorithms can be grouped into two categories: algorithms that are more or less efficient and ones that perform poorly. Along these lines, *FastGreedy* and *Leading Eigenvector* cannot identify the real community structure in any situation. The strength of the community structure clearly exerts a major influence on performances. As a matter of fact, performances generally decrease when the mixing parameter increases. Moreover, when the communities are well separated, all of the algorithms belonging to the first category easily uncover the community structure leading to high values of NMI. Differences are more pronounced when communities are more mixed. In this case, where communities are harder to detect, *Markov Cluster* and *Radetal* are not far from their limits. *InfoMap* outperforms all of the other algorithms under test. While *WalkTrap* and *Label Propagation* exhibit a good performance level, *Spinglass* and *Louvain* are one step behind. Note that the ranking deduced from the different measures of performance is quite comparable. Indeed, rank correlation values between any two performance measures are always higher than 0.95.

These results corroborate previous studies on the subject. In (Lancichinetti and Fortunato 2009), the authors conducted a comparative evaluation of twelve community-detection algorithms including *Radetal, Louvain*, *FastGreedy*, *InfoMap* and *Markov Cluster*. Performances have been computed using the NMI on two benchmark graphs with a known community structure (GN, LFR) and two random graphs (ER, CM) with no community structure. The results demonstrate that most methods perform rather well on the GN benchmark. But this is not the case when graphs are generated using the LFR. In particular, *FastGreedy* performs poorly. Its performances deteriorate rapidly when the network size increases and when community size decreases. *Radetal* and *Markov Cluster* are not very impressive either as their performances deteriorate with larger communities. As for the best algorithms, *InfoMap* ranks first followed by *Louvain*. When comparing these two, tests conducted on graphs with 100,000 nodes and community size ranging from 20 to 1000 have shown that *InfoMap* is not affected by these hard conditions, while it gets harder for *Louvain* to find the real communities. In addition, tests on random graphs are very informative. They should ideally lead to trivial solutions such as only one community or as many communities that there are nodes. Such is the case for *Radetal*, which always finds a single community with both benchmarks. *InfoMap* discovers no community on the ER benchmark but a few on the CM benchmark. On the contrary, *Markov Cluster* finds always as many communities as there are nodes. The remaining algorithms always find a few communities in these networks with no community structure. One of the major lessons of this work is that realistic benchmarks must be used for testing community-detection algorithms. In addition, the GN benchmark does not qualify to support the idea that new algorithms perform well as far as representing real-world networks adequately.

In (Orman, Labatut, and Cherifi 2011), eleven algorithms have been compared using the NMI. The LFR benchmark was tuned with parameters values typical of measurements on real-world networks. Results are in agreement with our previous observations concerning the ordering of the nine algorithms presented in table 11-1. *Infomap* outperforms the eleven algorithms under investigation. *WalkTrap*, *Label Propagation*, *Spinglass* and *Louvain* follow with satisfying performances, although not as good. *Markov Cluster* is very effective until the mixing proportion gets close to 0.4. Above this value, its performances decrease until a limit NMI value around 0.8. *Radetal* is very erratic and, *FastGreedy* and *Leading Eigenvector* are clearly outclassed. Besides the mixing coefficient, network size and average degree also affect the algorithms' performances but to a lesser extent. Performances worsen for *Spinglass*, *Leading Eigenvector*, *Louvain*, and *Radetal* when the network size increases, while the others are more insensitive to this parameter. To a certain extent, performances get better when the average degree increases. This may be due to an increase of the inter-community density that makes the community more visible. Undeniably, even if the intra-community links increase in the same proportion, the effect is distributed among many communities.

In (Coscia, Giannotti, and Pedreschi 2011), thirteen algorithms have been compared using a graph extracted from the ego network of one of the authors' Facebook profiles. In this small graph (261 nodes and 1,772 edges), communities are well separated. The authors had a perfect knowledge of the network's community structure. Thusly, the comparison relies on quality measures such as the modularity and the reverse conductance. Seven out of eight of the non-overlapping community-detection methods used in their evaluation are quite effective according to modularity. In fact, modularity ranges from 0.71 to 0.74. *Label Propagation, WalkTrap, and Edge Betweeness* tied for first, and were followed by *InfoMap*. Interestingly, *WalkTrap* favours a few bigger and denser communities, whereas *InfoMap* focuses on smaller and sparser ones. No solid conclusions can be drawn from this experiment. Of course, this network is not a very severe test and consequently not very discriminating for the algorithms. Moreover, testing the algorithms was not the primary goal of the authors in this work. They were more focused on classification according to the underlying community definition.

In (Orman, Labatut, and Cherifi 2012), the authors empirically tackle the relationship between performance measures and topological properties of the discovered community structure. The LFR benchmark is tuned in order to match the topological properties of three real-world networks (biological, Internet, communication). A comparative evaluation of eight algorithms is conducted using four performance measures (Purity, RI, ARI, NMI) and topological measures of the detected communities (embeddedness, community-size distribution, transitivity, scaled density, average distance, hub dominance). Table 11-2 summarizes some results for the algorithms previously evaluated. In extreme situations, the results are consistent. For example, *InfoMap* is still the best performing and *FastGreedy* is recognised as a substandard performer in both cases. Results are more mixed for the other algorithms. Thus, although the community structure identified by *Louvain* deviates significantly from the original, its NMI score is rather high. To summarise, there is no clear relation between these two types of evaluations that are rather complementary.

| Algorithm | d | Emb. | Dis. Size | Hub Dom. | Clust. | Sc. Dens. | NMI |
|---|---|---|---|---|---|---|---|
| *InfoMap* | Ok | Ok | Ok | Ok | Ok | Ok | 0.930 |
| *WalkTrap* | Ok | Ok | Ok | Ok | Ok | Ok | 0.865 |
| *Louvain* | | Ok | | | | | 0.735 |
| *Markov Cluster* | Ok | | Ok | | Ok | Ok | 0.881 |
| *FastGreedy* | | Ok | | | | | 0.588 |

**Table 11-2.** Qualitative comparison of the discovered communities' characteristics with the real community structure. The properties are average distance (d),

embeddedness (Emb.), community-size distribution (Dis. Size), hub dominance (Hub Dom.), transitivity (Clust.), and scaled density (Sc. Dens.). When both properties are akin, Ok is reported.

In order to investigate the realism level of synthetic benchmarks' effect on the performance of community-detection algorithms, eleven algorithms have been tested using LFR-CM, LFR-BA and LFR-EV (Orman, Labatut, and Cherifi 2013). The major trend is that performances deteriorate when the degree of the networks' realism increases. The highest performances are in general obtained when applied to the LFR-CM benchmark, whereas the lowest correspond to LFR-BA data. It turns out that *InfoMap* is the least sensible algorithm to network characteristic variations followed by *Spinglass* and *WalkTrap*. Figure 11-2 represents the best performing algorithms for each class of community definition. It shows that performances of *Spinglass* and *InfoMap* are insensitive to benchmark variations in a broad range of the mixing-proportion value. Differences appear when community discovery becomes a challenging task. On the contrary, *Radetal* and *Label Propagation* are affected almost in the whole range of mixing-proportion variation. The key lesson learned from this study is that even a slight variation of the network model can have important consequences on performances. It is therefore important to develop more realistic network models in order to have a clear view of the algorithms' efficiency.

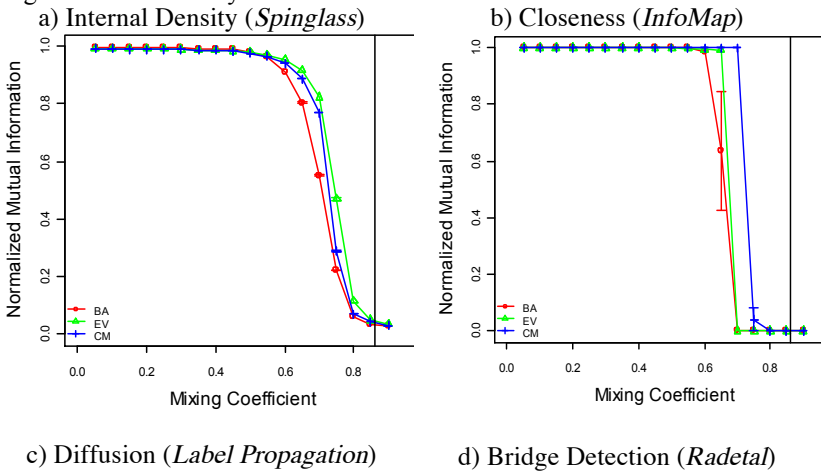

a) Internal Density (*Spinglass*)        b) Closeness (*InfoMap*)

c) Diffusion (*Label Propagation*)        d) Bridge Detection (*Radetal*)

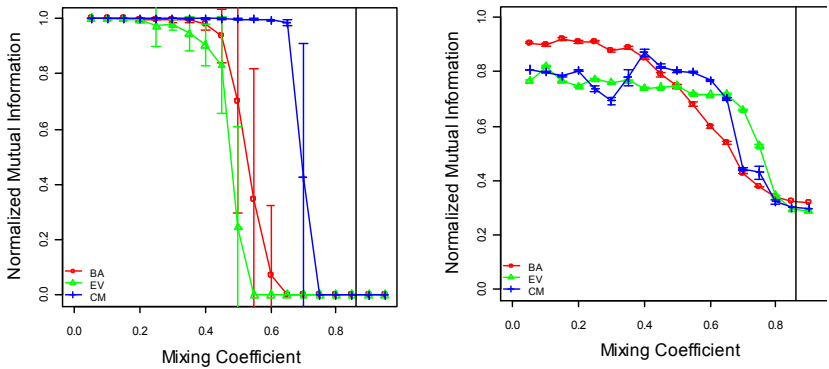

**Fig. 11-2.** Performances of the community-detection algorithms on the three benchmarks. Crosses are used for LFR-CM, dots for LFR-BA and triangles for LFR-EV. Each data point corresponds to an average over 25 generated networks with parameters values $n$ = 5000, $<k>$ = 30, $\gamma \approx 3$, $\beta$=2. The vertical lines at $\mu$ = 0.86 represent the average mixing limit above which communities stop being clearly defined.

Focusing on the various community-structure definitions, we will notice that three definition classes out of four have one of their members in the top four (*Spinglass* for internal density, *InfoMap* for closeness, and *Label Propagation* for diffusion). *Radetal* is one step behind the other categories. Unfortunately, as it is the only algorithm for the bridge-detection category, no serious conclusions can be drawn. Although it is somewhat risky to generalise from the study of a limited number of algorithms, it seems as if different definitions allow for the defining of equally efficient methods. To substantiate this intuition, it is necessary to conduct an extensive assessment on both synthetic and real-world data using a larger number of algorithms that must be selected among the top performers in each category.

# Conclusion

Initially focused on either microscopic or macroscopic properties, research on complex networks has shifted to mesoscopic properties of networks. While a great deal of work has been devoted to the community-detection issue, there are very few papers that analyse topological properties of the community structure in real-world networks. Strange as that might seem, most of the works are interested in discovering the community structure without really knowing what it may look like. Proposed solutions are usually based on a specific interpretation of an ill-defined concept of community. To demonstrate the effectiveness of their solution, small sized networks such as the *Zachary Karate Club* are used and (or) non-realistic synthetic benchmark graphs associated with traditional quality measures. This distorted view is aimed at pressing the point on the three complementary aspects of the community-detection issue: data, detection algorithms and objective evaluation. In order to ascertain a clear picture, progress must be made regarding these three directions.

Traditionally, in order to characterise real-world networks, research has been focused on macroscopic statistical properties to highlight their similarities. Empirical studies have demonstrated that numerous real-world networks share small-world and scale-free properties. It is now time to conduct a finer analysis of these networks to highlight their differences. Building a taxonomy of networks is one of the major steps in order to reach a consensus on ground-truth data to use for testing purposes. Furthermore, a taxonomy can have multiple benefits such as to shed light on networks' construction process. Mention may be made of two recent studies in this direction. In (Nicosia, De Domenico, and Latora 2013), node properties' values such as the degree, the average degree of the nearest neighbours, and the clustering coefficient are used to build time series generated by random walks. Based on these time series, a set of characteristic exponents is used to classify networks. More interestingly, the approach used in (Onnela et al. 2012) is based on the community structure of the networks. The authors introduce three mesoscopic response functions representing the evolution of the community structure according to a resolution parameter. The minimum value of this parameter forces each node into its own community, while the maximum value corresponds to a single community. The effective energy function is linked to the modularity. The two others, respectively denoted as "effective entropy" and "effective number of communities", are correspondingly a normalised measure of the entropy and the number of communities at a given scale. Based on these signatures, they build a taxonomy for 746 networks. Besides classifying networks in different categories, their approach allows for classifying networks within individual categories. These works can be exploited to define a set of real-world networks in order to test the algorithms in specific situations. They may also allow us to identify better what realistic properties according to the different network categories are. This idea is in line with the development of models for generating synthetic graphs with controlled properties. Indeed, there is room for improvement on models in order to provide an adequate description of real graphs with community structure. For instance, models whose transitivity can be adjusted might make for a significant refinement over existing solutions. Furthermore, knowledge about the topological properties of community structures must be developed in order to define appropriate models that can generate realistic networks. It is essential to progress in this area as tests on artificial data with controlled properties are needed to characterize detection algorithms. Undoubtedly, with real-world data, there is no guarantee that communities defined on a subjective basis are encoded in the structural information of the network. To date, even if there is some divergence on the way to organising community-detection algorithm categories, the situation gets clearer. The main difficulty is linked to the overlap between the definitions of community, which undermine the necessary consensus. To overcome this drawback, there is a need for someone to produce an extensive study on the complex connections between the algorithms and the definitions. Moreover, investigating if one definition is more relevant than the others or if these multiple views of the same problem are complementary is also essential. Finally, although traditional performance measures are used extensively, it appears very clearly that they cannot distinguish community structures with different topological properties. It is therefore necessary to propose alternative measures more sensitive to the community-structure variations in order to conduct more effective comparisons. Throughout this chapter we deliberately did not address the complexity issue. However, the analyst must not ignore this crucial information in order to choose an effective algorithm and a fair comparison should involve algorithms from the same class of complexity.